# Is one dimensional return map sufficient to describe the chaotic dynamics of a three dimensional system?

**Sayan Mukherjee [a], Sanjay Kumar Palit [b*], D K Bhattacharya [c]**

[a] Mathematics Department, Shivanath Shastri College, 23/49 Gariahat Road,
   Kolkata-700029, INDIA.
[b] Mathematics Department, Calcutta Institute of Engineering and Management,
   24/1A Chandi Ghosh Road, Kolkata-700040, INDIA.
[c] Rabindra Bharati University, Kolkata-700050, INDIA.

**Abstract.** Study of continuous dynamical system through Poincaré map is one of the most popular topics in nonlinear analysis. This is done by taking intersections of the orbit of flow by a hyper-plane parallel to one of the coordinate hyper-planes of co-dimension one. Naturally for a 3D-attractor, the Poincaré map gives rise to 2D points, which can describe the dynamics of the attractor properly. In a very special case, sometimes these 2D points are considered as their 1D-projections to obtain a 1D map. However, this is an artificial way of reducing the 2D map by dropping one of the variables. Sometimes it is found that the two coordinates of the points on the Poincaré section are functionally related. This also reduces the 2D Poincaré map to a 1D map. This reduction is natural, and not artificial as mentioned above. In the present study, this issue is being highlighted. In fact, we find out some examples, which show that even this natural reduction of the 2D Poincaré map is not always justified, because the resultant 1D map may fail to generate the original dynamics. This proves that to describe the dynamics of the 3D chaotic attractor, the minimum dimension of the Poincaré map must be two, in general.

Keywords: Chaotic attractor, Poincaré section, Poincaré map.

## 1. Introduction

The dynamics of a high dimensional flow in the corresponding phase space [1-6] is understood conventionally by observing the dynamics induced by the flow on a particular section of the phase space. The chosen section, called the Poincaré section [7] helps in visualizing the underlying dynamics. The successive intersections of the flow with the section produce a discrete mapping known as the Poincaré map or the return map [8-15]. In fact, Poincaré map [8-15] is a standard tool to study qualitative properties of a dynamical system, most prominently the asymptotic stability of periodic or almost periodic orbits. Such an orbit $O$ (called the reference orbit) and its neighbouring orbits transverse the Poincaré section $S$, which results in some points on $S$. The key aspects of the Poincaré map [8-15] is that it essentially describes how such points on the Poincaré section $S$ get mapped back onto $S$ by the flow. The major advantages in this approach include the analysis of the long-term behavior of the flow close to *O* through the derivative of the Poincaré map [8-15] at the point of intersection of S and *O* available just after one revolution of *O*, and the reduction of the dimension of the problem by one, since the Poincaré map [8-15] is defined on *S* and neglects the "trivial" direction of the flow perpendicular to the surface. However, it is a common practice to take return map as a one dimensional one [14]. The familiar Rossler flow and its Poincaré section is one such example. The flow wraps around the *Z*-axis, so a good choice for a Poincaré section is a plane passing through the *Z*-axis. A sequence of such Poincaré sections placed radially at increasing angles with respect to the *X*-axis illustrates the 'stretch and fold' action of the Rossler flow. Comparing the different *Z*-axis scales of these sections a series of snapshot of the flow is observed. A line segment traversing the width of the attractor starts very close to the *XY*-plane and after the stretching followed by the folding, the folded segment returns close to the *XY*-plane strongly compressed. Once a particular Poincaré section is picked-up, we can also exhibit the return map. The cases are examples of nice one-to-one return maps. However, they appear multimodal and non-invertible artifacts of projection of a 2*D* return map $(R_n, z_n) \to (R_{n+1}, z_{n+1})$ onto a 1*D* subspace $R_n \to R_{n+1}$. The above examples illustrate why a Poincaré section gives a more informative snapshot of the flow than the full flow portrait. For example, while the full flow portrait of the Rossler flow gives us no sense of the thickness of the attractor, we see that though the Poincaré sections and even the return map is 2*D*  2*D*, the flow contraction is so strong that for all practical purposes it renders the return map 1-dimensional.

[*] Corresponding author. Tel.: +919831004544.
   *E-mail address*: sanjaypalit@ yahoo.co.in

In this connection, we may remember Lorenz map, which is also a one dimensional map. But the difference is that this one dimensional map does not follow as a particular case of a 2*D* Poincaré map. In fact, what Lorenz observed is that - the trajectory apparently leaves one spiral only after exceeding some critical distance from the center [3]. Moreover, the extent to which this distance is exceeded appears to determine the point at which the next spiral enters; this in turn seems to determine the number of circuits to be executed before changing spirals again. It therefore seems that some single feature of a given circuit should predict the same feature of the following circuit. The "single feature" that he focuses on is $z_n$ the $n^{th}$ local maximum of $z(t)$, $z(t)$ being one of the solution components of the Lorenz dynamical system. Thus, Lorenz could explain the dynamics by the one dimensional map $z_{n+1} = f(z_n)$ for some suitable *f*. However this simpler approach works only if the attractor is very "flat," i.e., close to two-dimensional, as the Lorenz attractor is (fractal dimension 2.03) [3]. Hence, such 1*D* map is never possible in a general case of attractor whose fractal dimension is far from two. In this article, we address this issue and establish that consideration of 2*D* maps as the descriptor of a 3*D* dynamical system as a remedy to this problem.

## 2. Materials and Methods

*2.1. Poincaré map of three dimensional attractors of some dynamical systems*

Let $\dot{\vec{x}} = f(\vec{x}), \vec{x} \in \mathbb{R}^n$ be an autonomous differential equation in *n* dimensional Euclidean space. The basic idea of Poincaré was to consider only the flow's piercings of a $n-1$ dimensional "transverse surface", called Poincaré section. The intersections are points and all such intersections form the so-called Poincaré map. The choice of the Poincaré section [7] is altogether arbitrary. It is rarely possible to choose a single section that cuts across all trajectories.

In a more detail, suppose a trajectory on a three dimensional attractor meets the Poincaré section [7] (taken as a plane parallel to one of the coordinate planes) at a point $(u_1, v_1)$. Let the next points of intersection of the trajectory of the attractor with that Poincaré plane are

$(u_2, v_2), (u_3, v_3)$, and so on. If there exist two functions $P_u, P_v$ such that $u_{n+1} = P_u(u_n, v_n), v_{n+1} = P_v(u_n, v_n), n = 1, 2, 3, ...$, then the intersection of these two three dimensional surfaces is a two dimensional curve [16]. This is called the two dimensional Poincaré map [8-15] corresponding to the Poincaré section [7] passing through the points $(u_n, v_n), n = 1, 2, 3, ....$.

*2.2. Reduction of two dimensional Poincaré map to one dimensional one*

As there is no fixed rule to take the plane of section parallel to the coordinate planes to get the Poincaré map, so there is a possibility that two dimensional Poincaré map [8-15] may reduce to a one dimensional one under a suitable choice of the plane of section. In fact, this occurs if the points $(u_n, v_n)$ on the Poincaré section [7] become functionally related as $v_n = F(u_n)$, or $u_n = F_1(v_n)$. In that case we get $u_{n+1} = P_u(u_n, F(u_n)) = G(u_n)$. This is a one dimensional map. We may check graphically, whether $v_n = F(u_n)$. So we draw a two dimensional graph with $u_i$'s on the X-axis and $v_i$'s on the Y-axis and check, whether corresponding to each $u_i$, there is just one $v_i$ or not. In case, it is just one, it is proved that $v_n = F(u_n), \forall n$ and the map reduces to a one dimensional one. Otherwise, the two dimensional map is not reducible to a one dimensional one.

**3. Results and Discussion**

*3.1. Poincaré map of the three dimensional attractor of Neuro-dynamical system*

We consider a known Neuro-dynamical system [17] model described as follows:

$$\frac{dx}{dt} = \left[1 + \exp\{-s_1(w_{21}y + w_{31}z - \eta_1)\}\right]^{-1} - r_1 x$$
$$\frac{dy}{dt} = \left[1 + \exp\{-s_2(x - \eta_2)\}\right]^{-1} - r_2 y \tag{1}$$
$$\frac{dz}{dt} = \left[1 + \exp\{-s_3(x - \eta_3)\}\right]^{-1} - r_3 z$$

with the initial condition $x(1) = 0.8, y(1) = 0.5, z(1) = 0.1$ and the parameters given by $w_{21} = 1, w_{31} = -6.2, r_1 = 0.62, r_2 = 0.42, r_3 = 0.1, s_1 = 7, s_2 = 7, s_3 = 13, \eta_1 = 0.5, \eta_2 = 0.3, \eta_3 = 0.7$.
This choice of parameters gives rise to a chaotic attractor as shown by fig. 1.

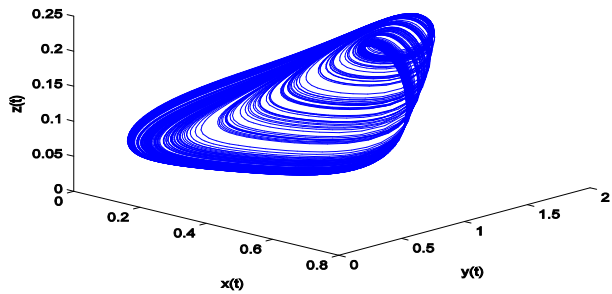

**Fig. 1. Three dimensional chaotic attractor of Neuro-dynamical system.**

To find Poincaré map [8-15], we first cut this attractor by a plane $z = 0.0963$, keeping $(x, y)$ varying of its own. We chose 500 such points $(x, y)$, which lie on the aforesaid plane, known as Poincaré plane of section. Next we denote these points by $\{(u_k, v_k)\}_{k=1}^{1000}$. We find that the locus of these points is a curve as shown by fig. 2.

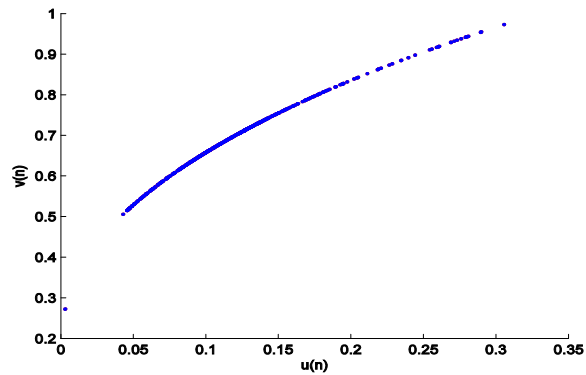

**Fig. 2. The U – V curve of the Neuro-dynamical system.**

To understand that this curve is represented by a two-dimensional map, we are to prove that it is the intersection of two surfaces $u_{n+1} = f(u_n, v_n), v_{n+1} = g(u_n, v_n), n = 1, 2, 3, ..$ for some choice of functions $f$ and $g$. For this purpose, we plot $u_{n+1}$ against $(u_n, v_n)$ and $v_{n+1}$ against $(u_n, v_n)$. These are shown by fig. 3a and fig. 3b respectively.

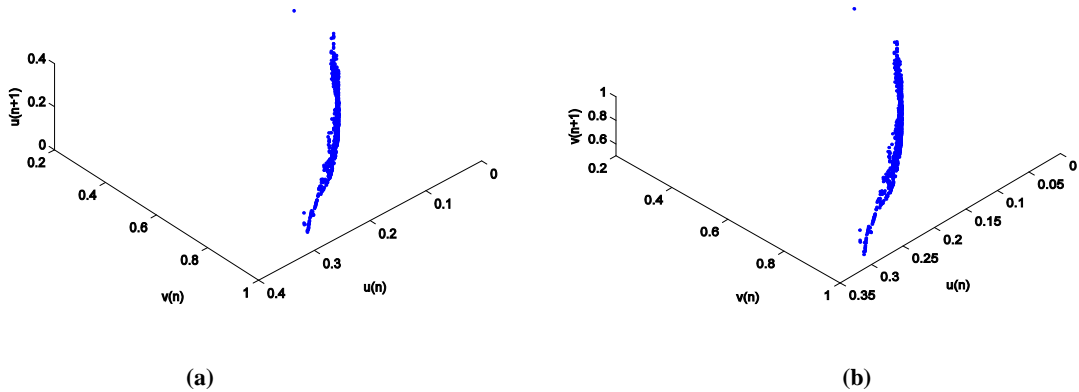

**Fig. 3. (a) Plot of $u_{n+1}$ against $(u_n, v_n)$, (b) plot of $v_{n+1}$ against $(u_n, v_n)$.**

To find the nature of the surfaces given by fig.3 (a) and fig. 3 (b), we try for 'curve of best fit' for them. These are given by fig.4 (a) and fig. 4 (b) respectively.

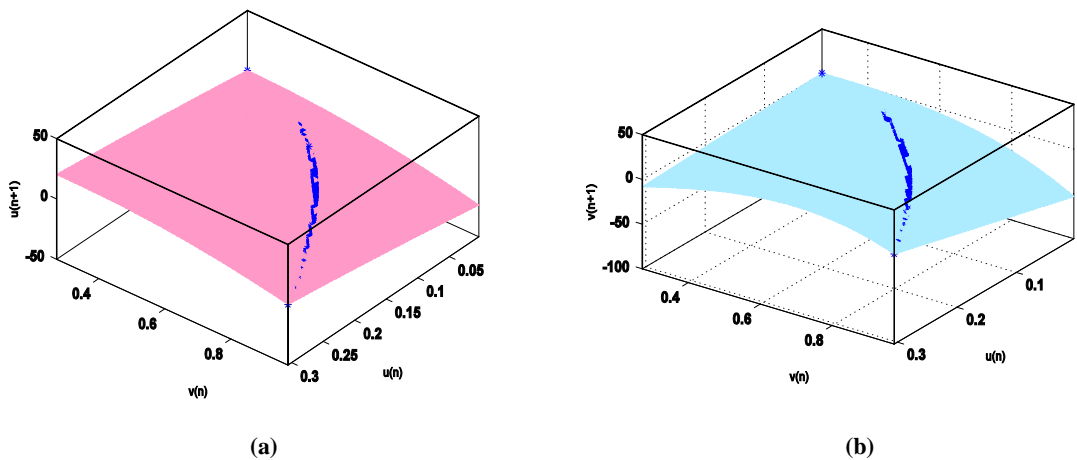

**Fig. 4. (a) Fitted surface: $u_{n+1} = f(u_n, v_n)$, (b) fitted surface : $v_{n+1} = g(u_n, v_n)$.**

It is seen that the intersection of these two surfaces given by fig. 5a resembles the original U -V curve of fig. 2. This proves that it is a two dimensional curve.

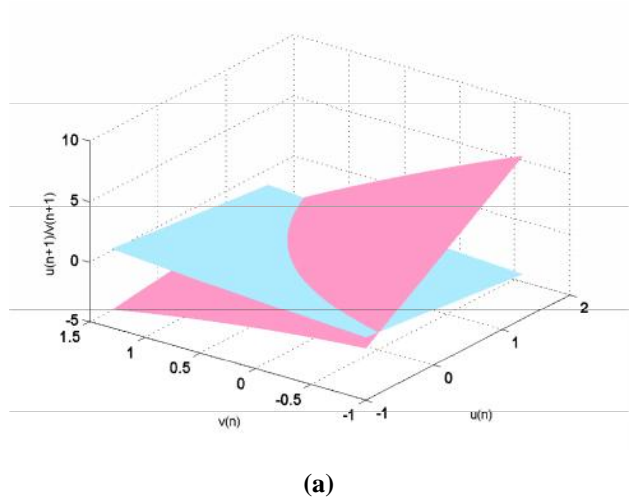

**(a)**

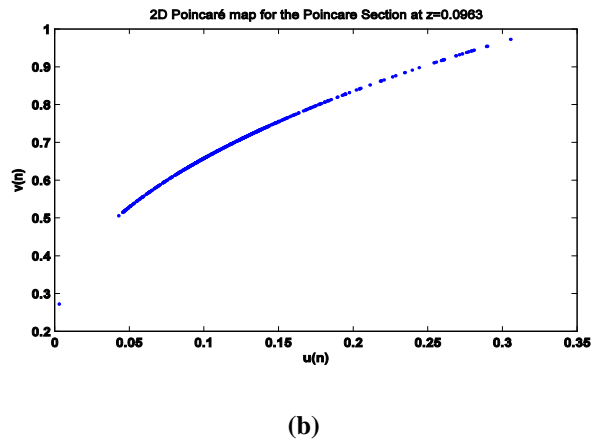

**(b)**

**Fig. 5. (a) 3D curve made by the intersection of the surfaces given by fig. 4a and fig. 4b,
(b) 2D projection of 3D curve of intersection of the surfaces.**

It is also found that the surfaces in fig. 4a and fig. 4b take the form
$f(u_n, v_n) = -2.267 + 73.31u_n + 19.79v_n - 37.45u_n^2 + 17.26u_n v_n - 43.28v_n^2$
and $g(u_n, v_n) = 1.665 - 135.9u_n - 18.12v_n + 416 u_n v_n + 92.62 v_n^2 - 96.67u_n v_n^2 - 136.1 v_n^3$
respectively.

This shows that both the functions $f$ and $g$ represents surfaces and the present Poincaré map [8-15] is almost like a two dimensional map, which is a polynomial in both the variables.

As it is known that such two dimensional maps, polynomial in both the variables are capable of showing proper chaotic behavior under suitable choice of parameters [18,19], so it is concluded that such two dimensional Poincaré maps [8-15] exhibit chaotic nature of the original attractor.

*3.2. Irreducibility of 2D Poincaré map for Neuro-dynamical system*

Apparently, the 2D projection of the 3D curve of intersection of the surfaces given by fig.5b seems to be a two dimensional curve and so one may try to fit a suitable curve to it. But this is not a correct approach and one requires checking vividly before coming into the conclusion that the curve is two dimensional. This is because, if a curve is fitted to that apparently looking 2D curve, it means that the 2D Poincaré map is reducible to 1D Poincaré map.

Thus, when the question arises whether a 2D Poincaré map is reducible to a 1D Poincaré map or not, it is necessary to check whether for each $u_n$ there exists exactly one $v_n$ or not. For this purpose, fig.5b is zoomed to obtain fig.6 from which it is seen that for $u_n = 0.1291$, there are two different values of $v_n$ given by $v_n = 0.7287, 0.7288$. Hence, $v_n \neq f(u_n)$. Hence the 2D Poincaré map of Neuro-dynamical system for the Poincaré section $z = 0.0963$ is not reducible to 1D Poincaré map.

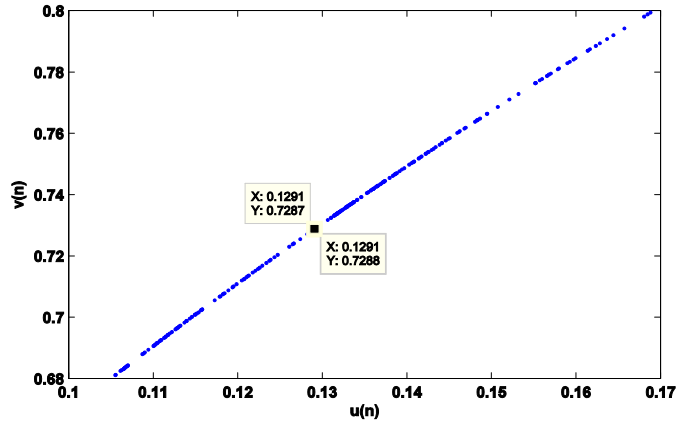

**Fig.6. 2D projection of the 2D Poincare map of Neuro-dynamical system (zoomed version).**

*3.3. Poincaré map of three dimensional attractors of Lorenz system*

We consider Lorenz dynamical system [20] given by Eq. (2).

$$\frac{dx}{dt} = s\,(y - x)$$
$$\frac{dy}{dt} = r\,x - y - x\,z \quad (2)$$
$$\frac{dz}{dt} = -b\,z + xy$$

with initial conditions: $x(1)=8, y(1)=9, z(1)=25$.

It gives the following chaotic attractor under the parameters $s = 10, r = 28, b = \frac{8}{3}$.

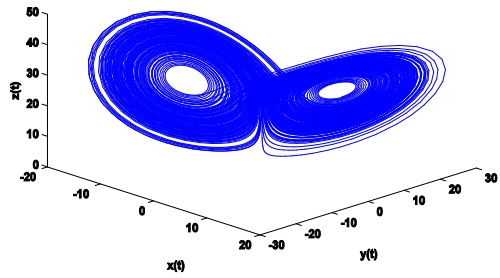

**Fig. 7. Three dimensional chaotic attractor of Lorenz system.**

To find Poincaré map [8-15], we first cut this attractor by a plane $z = 16$, keeping $(x, y)$ varying of its own. Proceeding as above, we get the U – V curve as given in fig. 8.

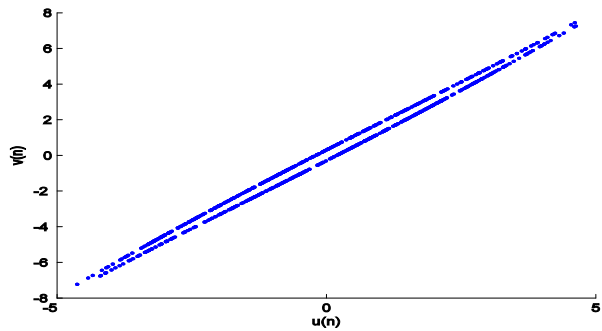

**Fig. 8. The U – V curve of Lorenz dynamical system.**

To understand that this curve represents a two dimensional map, we find two surfaces $u_{n+1} = f(u_n, v_n), v_{n+1} = g(u_n, v_n), n = 1, 2, 3,..$ under suitable functions $f$ and $g$. For this purpose, we plot $u_{n+1}$ against $(u_n, v_n)$ and $v_{n+1}$ against $(u_n, v_n)$ to obtain two surfaces as given by fig. 9a and fig. 9b respectively.

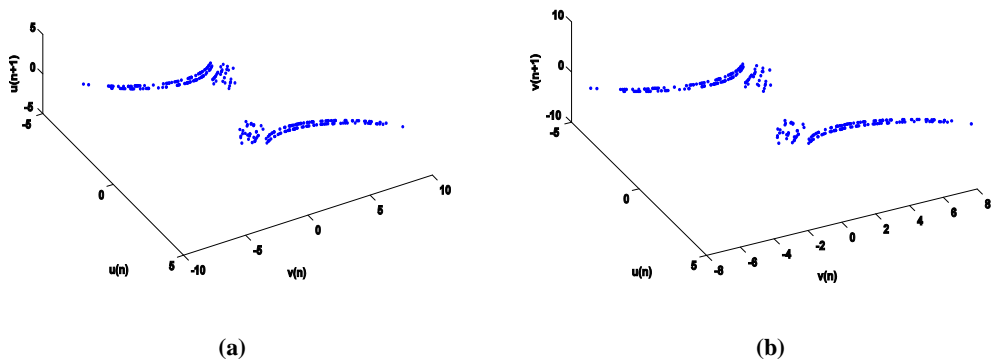

(a)  (b)

**Fig. 9. (a) Plot of $u_{n+1}$ against $(u_n, v_n)$, (b) Plot of $v_{n+1}$ against $(u_n, v_n)$.**

To understand the nature of the surfaces given by fig. 9a and fig. 9b, we try for 'curve of best fit' for them. These are given by fig. 10a and fig. 10b respectively.

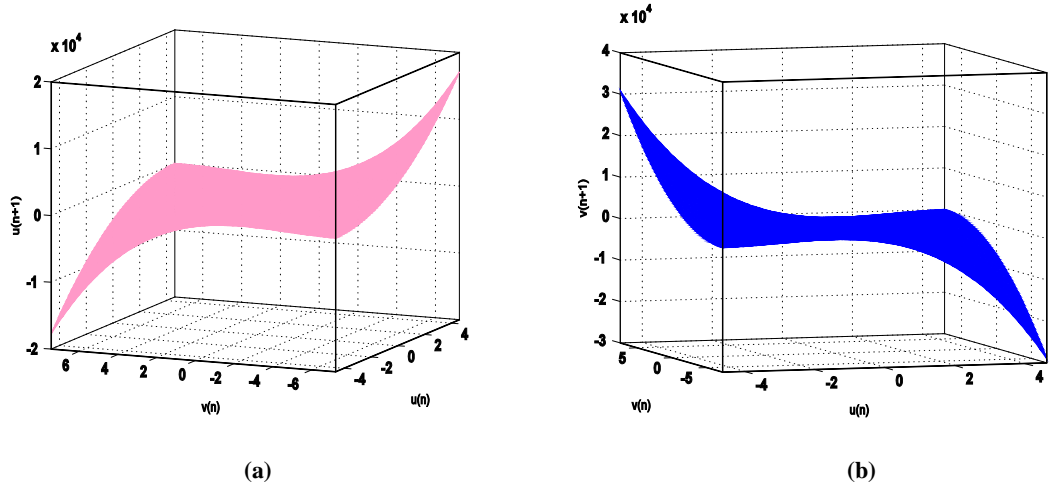

(a)                            (b)

**Fig. 10. (a) Fitted surface $u_{n+1} = f(u_n, v_n)$, (b) fitted surface $v_{n+1} = g(u_n, v_n)$.**

Fig. 11 shows the 2D projection of the intersection of the two surfaces drawn in 3D as in case of Neuro-dynamical system [section.3.1]. It is found that this 2D curve resembles the original U – V curve.

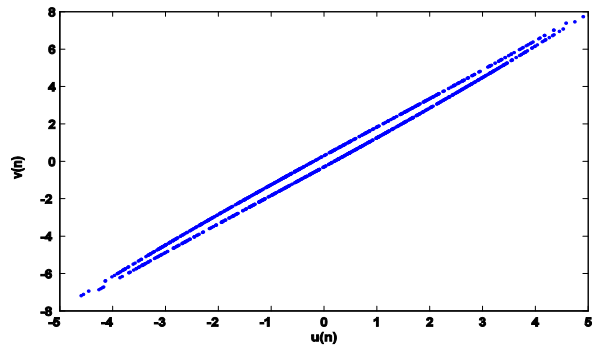

**Fig. 11. 2D projection of 3D curve of intersection of the surfaces.**

It is also found that surfaces in fig. 10a and fig. 10b are taking the form of
$$f(u_n, v_n) = 0.3196 + 0.186 u_n + 0.4386 v_n - 36.64 u_n^2 + 71.51 u_n v_n - 34.95 v_n^2 - 654.5 u_n^2 v_n + 1302 u_n v_n^2 - 646.5 v_n^3$$
and
$$g(u_n, v_n) = 0.1795 + 2.056 u_n - 1.021 v_n - 3.388 u_n^2 + 4.325 u_n v_n - 1.386 v_n^2 - 48.79 u_n^3 + 62.41 u_n^2 v_n - 19.94 u_n v_n^2$$
respectively.

It again appears that both the functions $f$ and $g$ represent surfaces. As explained above, it is capable of showing proper chaotic behavior of the original attractor under suitable choice of parameters.

*3.4. Irreducibility of 2D Poincaré map for Lorenz system*

It is evident from fig.11 that the 2D projection of the intersection of the two surfaces in 3D for the Poincaré section $z=16$ of the Lorenz system is not at all a 2D curve and hence it cannot be reduced to a 1D map. To avoid any confusion in this regard, fig.11 is zoomed to obtain fig.12 from where it is seen that for $u_n = -1.784$, there exist two different value of $v_n$ given by $v_n = -2.497, -3.018$. Hence, $v_n \neq f(u_n)$. Thus 2D Poincaré map of Lorenz system for the Poincaré section $z = 16$, is not reducible to 1D Poincaré map.

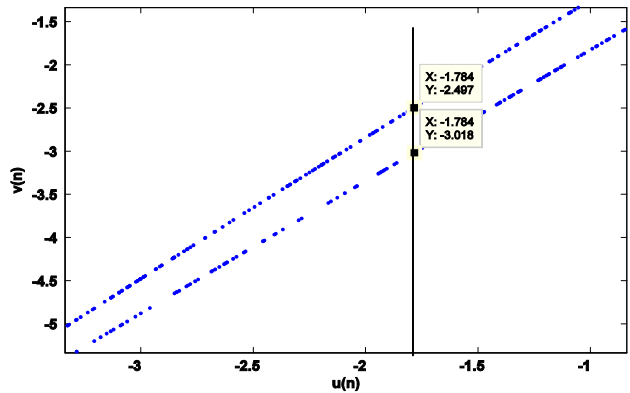

**Fig.12. 2D projection of the 2D Poincare map of Lorenz system (zoomed version).**

*3.5. Reducible two dimensional Poincaré map*

Reducibility of a two dimensional Poincaré map [8-15] lies in suitable manipulation in the choice of plane of section in some cases but this is not true in general. For example, we now take z=30.5 as the plane of section of the same Lorenz attractor as given by fig. 7. Proceeding similarly as earlier (section 3.1 and section 3.3), the 2D projection of the 3D curve of intersection of surfaces is obtained (fig. 13b). Now the 2D map reduces to a 1D Poincaré map (which is shown in fig.13c).

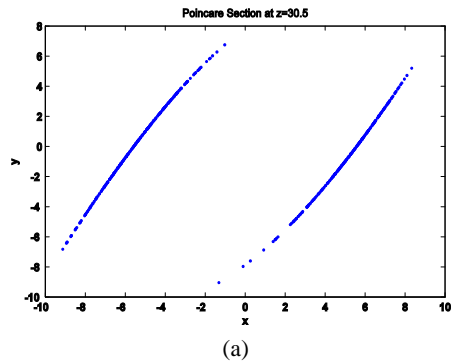

(a)

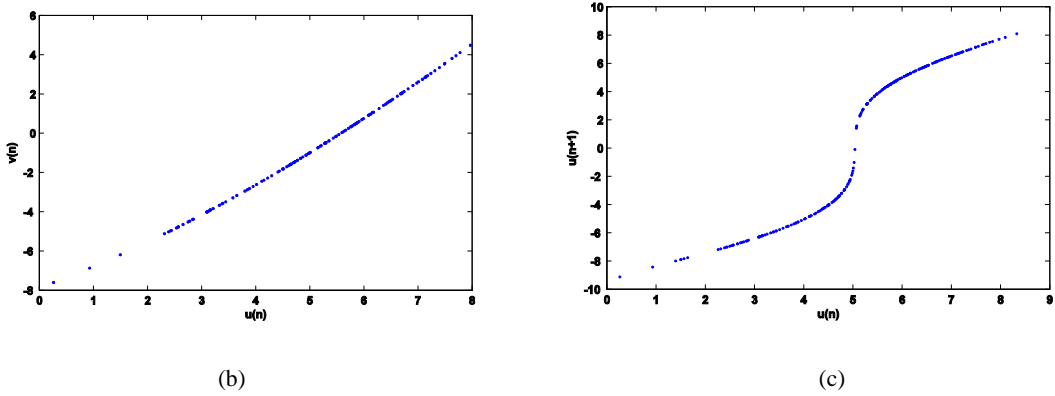

(b)                                      (c)

**Fig. 13. (a) The Poincaré section of Lorentz attractor, (b) 2D Poincaré map,
(c) 1-D Poincaré map [One-sided Poincaré section].**

In order to be sure that corresponding to each $u_n$ there exists exactly one $v_n$, the alternative way is to verify that all of $u_1, u_2, u_3, ........, u_{N+1}$ (where $N$ is the number of points in the 1D map, say) are distinct. Let

$$d(i, j) = \begin{cases} 1, & |u_i - u_j| = 0 \\ 0, & |u_i - u_j| \neq 0. \end{cases}, \; i, j = 1, 2, 3, ......, N+1.$$

The matrix $d$ is presented graphically in fig.14, where the black dots represent '1' and the white dots represent '0'.

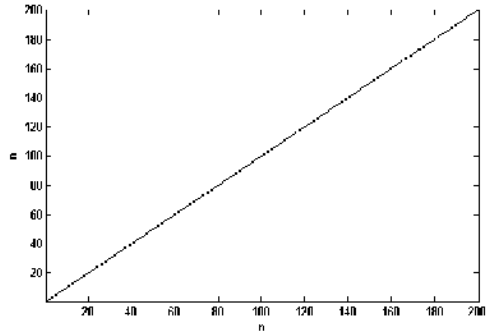

**Fig14. Matrix Plot of the matrix $d$.**

It is observed from fig.14 that the black dots are located only along the main diagonal. This proves that $|u_i - u_j| \neq 0 \; \forall i, j = 1, 2, 3, ......, N+1, i \neq j$. It means that no two values of $u_n$ coincides and thus neither it is possible to get more than one $v_n$ for any $u_n$. Thus fig.13c represents a 1D map of the Lorenz system.

As the characteristics of the orbits of a three dimensional differential equation is fully shown by the corresponding 2D Poincaré section, so it is natural to expect that the chaotic attractor of the 3D Lorenz system is properly exhibited in this 2D section. As our section is a very, very particular choice of the two dimensional Poincare section in order to get one dimensional map, so the full chaotic nature of the 2D map, in other words the strict positivity of the Lyapunov exponent calculated there from, may not be followed in Toto. Naturally, chaotic orbits near the stable attractor may be observed in such a case. Actually Lyapunov exponent calculated from our data set (0.0023) is no doubt very near to zero but it is never negative. So, there is a chaotic orbit but not as general as is expected.

**4. Conclusions**

Occurrence of one dimensional Poincaré map is a rare phenomenon as it may be obtained by a very special choice of Poincaré section. In fact, one dimensional Poincaré map [8-15] is unable to explain the chaotic dynamics of the three dimensional attractor in most of the cases. Actually for most of the three dimensional attractors, one cannot find out one dimensional Poincaré map [8-15]. Even when the one dimensional Poincaré map is obtained for some dynamical systems, it is not workable, as it may not describe the proper chaotic dynamics of the attractor. However, the present study confirms that it is always possible to obtain two dimensional Poincaré map [8-15] for any three dimensional dynamical system and in general it is not reducible to a one dimensional Poincaré map. Since such 2D Poincaré map [8-15] is capable of studying the chaotic behavior of the system, so it may be concluded that two dimensional Poincaré maps are the maps of minimum dimension, which are capable to explain the dynamics of three dimensional attractors.